\documentclass[]{article}
\usepackage{proceed2e}
\usepackage[protrusion=true,spacing=true]{microtype}
\usepackage{amsmath,amssymb,amsfonts,amsthm}
\usepackage{algorithmic,algorithm}
\usepackage{graphicx,cite}
\usepackage[absolute,overlay]{textpos}

\newtheoremstyle{mystyle}
  {}			% Space above
  {0pt}		% Space below
  {\itshape}	% Body font
  {}			% Indent amount
  {\bfseries}	% Theorem head font
  {.}			% Punctuation after theorem head
  { }			% Space after theorem head, ' ', or \newline
  {}			% Theorem head spec (can be left empty, meaning `normal')
\theoremstyle{definition}
\newtheorem{_definition}{Definition}
\newtheorem{_proposition}{Proposition}

\newcommand{\td}{\Delta}
\newcommand{\TD}{\mathbb{D}}
\newcommand{\pr}{\text{Pr}}

\title{A Game-Theoretic Model and Best-Response Learning \\ Method for Ad Hoc Coordination in Multiagent Systems}

\author{
	{\bf Stefano V. Albrecht} \\
	School of Informatics \\
	University of Edinburgh \\
	Edinburgh EH8 9AB, UK \\
	\texttt{s.v.albrecht@sms.ed.ac.uk}
	\And
	{\bf Subramanian Ramamoorthy} \\
	School of Informatics \\
	University of Edinburgh \\
	Edinburgh EH8 9AB, UK \\
	\texttt{s.ramamoorthy@ed.ac.uk}
}

\begin{document}

	\maketitle

	\begin{textblock}{12.7}(1.97,0.8)
		\small \textbf{Technical Report:} The University of Edinburgh, February 2013. \hfill Last revised: January 2014
	\end{textblock}

	% ABSTRACT
	\begin{abstract}
	\vspace{-5pt}
The \emph{ad hoc coordination} problem is to design an autonomous agent which is able to achieve optimal flexibility and efficiency in a multiagent system with no mechanisms for prior coordination. We conceptualise this problem formally using a game-theoretic model, called the \emph{stochastic~Bayesian~game}, in which the behaviour of a player is determined by its private information, or \emph{type}. Based on this model, we derive a solution, called \emph{Harsanyi-Bellman Ad Hoc Coordination}~(HBA), which utilises the concept of Bayesian Nash equilibrium in a planning procedure to find optimal actions in the sense of Bellman optimal control. We evaluate HBA in a multiagent logistics domain called \emph{level-based foraging}, showing that it achieves higher flexibility and efficiency than several alternative algorithms. We also report on a human-machine experiment at a public science exhibition in which the human participants played repeated Prisoner's Dilemma and Rock-Paper-Scissors against HBA and alternative algorithms, showing that HBA achieves equal efficiency and a significantly higher welfare and winning rate.
	\vspace{5pt}
	\end{abstract}

	% INTRODUCTION
	\section{Introduction} \label{sec:intro}

We are concerned with the \emph{ad hoc coordination} problem, in which the goal is to design an autonomous agent, called the \emph{ad hoc agent}, which is able to achieve optimal flexibility and efficiency in a multiagent system that admits no prior coordination between the ad hoc agent and the other agents. \emph{Flexibility} describes the ad hoc agent's ability to solve its task with a variety of other agents in the system. \emph{Efficiency} is the relation between the ad hoc agent's payoffs and time needed to solve the task. \emph{No prior coordination} means that the ad hoc agent does not know ahead of time who the other agents are and how they behave. In particular, there are no prior agreements on information sharing, communication and action protocols, standards, etc.

This problem is motivated by the fact that there is a growing number of agents, both robotic and virtual, which are employed in an increasing number of areas. Given that a primary goal in agents research is to increase the autonomy and thus lifetime of agents, it can be expected that agents based on different technologies may have to interact in nontrivial ways, without knowing a priori who the other agents are. This motivates both the notion of flexibility, since the other agents could be based on any kind of technology, and efficiency, since there may be no time for long learning periods, especially if interactions are sparse. Human-machine interaction problems (e.g. robots used in rescue scenarios or software agents used in trading markets) can be viewed as a special case of ad hoc coordination, since humans have extremely variable behaviour (flexibility) and expect agents to be able to interact quickly (efficiency), while there may be no prior description of the human's behaviour (no prior coordination).

There have been several attempts to address ad hoc coordination in multiagent systems, e.g. \cite{bm2005,dhb2006,skkr2010}. While all of these works are relevant to ad hoc coordination, the assumptions made by the solutions therein imply that they only address certain aspects of the larger problem. For example, in \cite{bm2005,dhb2006} it is assumed that all agents follow pre-specified plans which include roles and synchronised action sequences for each role, and in \cite{sk2010,skr2010,bsk2011,as2012} it is assumed that the other agents' behaviours are fixed and known, and that all agents have common payoffs. We also note that the problem descriptions in these works are of a procedural nature, associated with the specific tasks considered therein. Therefore, there is a need for a formal model of the ad hoc coordination problem, general enough to accommodate a wide spectrum of problems.

A related problem is known in game theory as the \emph{incomplete information game}. Therein, each player has some private information relevant to its decision making of which the other players are not aware, which is what relates the incomplete information game to the ad hoc coordination problem. \cite{h1967} introduced \emph{Bayesian games} in which the private information of a player is abstractly represented by its \emph{type}, admitting a solution in the form of the \emph{Bayesian Nash equilibrium}. Since then, there have been several works on learning in Bayesian games, e.g. \cite{j1991,kl1993,dfl2004}. While the notion of private information is useful to describe the ad hoc coordination problem, the learning processes and solutions studied therein are not directly applicable, since the focus has traditionally been on equilibrium considerations but not on efficiency. On the other hand, much work in multiagent systems has focused on efficiency, whilst often making central assumptions about the other agent's behaviours \cite{ar2012}. Therefore, it is natural to ask if these fields can be combined to address ad hoc coordination in a useful way.

Inspired by this question, we model the problem using a game-theoretic construct called the \emph{stochastic Bayesian game}, in which a player's behaviour is determined by its \emph{type}. Based on this model, we give formal definitions of flexibility and efficiency, and we define ad hoc coordination as the problem of optimising flexibility and efficiency, subject to the constraint that the ad hoc agent is unaware of the players' \emph{type spaces}, and hence the rules by which their types are assigned. Our model allows for both the definition of Bayesian Nash equilibrium and, since it satisfies the Markov property, the definition of \emph{Bellman optimal control} \cite{b1957}, a key result in intelligent agents. We combine these two concepts to obtain a solution which we call \emph{Harsanyi-Bellman Ad Hoc Coordination} (HBA). HBA does not rely on a central assumption about the other agents' behaviours. Instead, it allows for the specification of multiple such assumptions which are provided to HBA as a set of \emph{user-defined types}, each corresponding to a different hypothesis of how an agent might behave. Based on the agents' observed actions, HBA computes probability distributions over the user-defined types, called \emph{posteriors}, and utilises them in a planning procedure to find optimal actions.

HBA has a number of useful features with respect to ad hoc coordination. The fact that the user-defined types may encapsulate any kind of behaviour means that HBA can potentially deal with a variety of different agents, including agents which maintain beliefs about the behaviour of the HBA agent, or any other type of recursive reasoning. We show this in a human-machine experiment conducted at a public science exhibition, in which HBA was able to manipulate the beliefs of humans in repeated Prisoner's Dilemma such that both ended up cooperating, thus maximising its efficiency. HBA also supports the possibility that agents may switch between different behaviours. We address this by introducing \emph{temporally reweighted posteriors} which allow HBA to quickly recognise changed types. In our human-machine experiment, this allowed HBA to achieve a significantly higher winning rate in Rock-Paper-Scissors than the human participants and an alternative algorithm.

A central feature of HBA is that it can use the types to plan in the entire state space of the problem (including unseen states) provided that the posteriors and user-defined types are reasonably accurate. To accommodate the case in which none of the user-defined types accurately describe an agent's behaviour, HBA is able to include methods for opponent modelling. We propose an opponent modelling method, called \emph{conceptual type}, which can be viewed as a kind of type that specifies the \emph{conceptualisation} underlying a behaviour, rather than specifying the behaviour directly. The conceptualisation is combined with the observed actions of an agent to generalise its actions to unseen states and improve accuracy in rarely visited states. We demonstrate these features in a multiagent logistics domain called \emph{level-based foraging}, in which HBA is able to achieve significantly higher flexibility and efficiency than three alternative algorithms (JAL \cite{cb1998}, CJAL \cite{bs2007}, WoLF-PHC \cite{bv2002}), using just a few user-defined types.

	% DEFINING AD HOC COORDINATION
	\section{Defining Ad Hoc Coordination} \label{sec:def}

		% STOCHASTIC BAYESIAN GAMES
		\subsection{Stochastic Bayesian Games} \label{sec:sbg}

As discussed earlier, ad hoc coordination can be defined based on the notion of private information in Bayesian games. However, in their original form \cite{h1967}, Bayesian games are not descriptive enough to allow us to model the kinds of problems we are interested in, as they do neither include states nor time. Therefore, we combine Bayesian games with the concept of \emph{stochastic games} \cite{s1953} to obtain a more descriptive model which we call \emph{stochastic Bayesian game}:\footnote{A related model are I-POMDP, in which agents face incomplete information with respect to the state of the world and the behaviour of other agents \cite{gd2005}. However, I-POMDP are extremely complex and their solution methods are infeasible in most problems.}

\begin{_definition}
	A \emph{stochastic Bayesian game} (SBG) consists of:

	\vspace{-10pt}

	\begin{itemize}
		\setlength{\itemsep}{0pt}
		\item discrete state space $S$ with initial state $s^0 \in S$ and terminal states $\bar{S} \subset S$
		\item players $N = \left\{ 1,...,n \right\}$ and for each $i \in N$:
		\vspace{-5pt}
		\begin{itemize}
			\setlength{\itemsep}{0pt}
			\item set of actions $A_i$ (where $A = A_1 \times ... \times A_n$)
			\item type space $\Theta_i$ (where $\Theta = \Theta_1 \times ... \times \Theta_n$)
			\item payoff function $u_i : S \times A \times \Theta_i \rightarrow \mathbb{R}$
			\item strategy $\pi_i : \mathbb{H} \times A_i \times \Theta_i \rightarrow [0,1]$
		\end{itemize}
		\item state transition function $T : S \times A \times S \rightarrow [0,1]$
		\item type distribution $\td : \mathbb{N}_0 \times \Theta \rightarrow [0,1]$
	\end{itemize}

	\vspace{-10pt}

	$\mathbb{H}$ contains all \emph{histories} $H^t = \langle s^0,a^0,s^1,a^1,...,s^t \rangle$ with $t \geq 0$, $(s^\tau,a^\tau) \in S \times A$ for $0 \leq \tau < t$, and $s^t \in S$.
\end{_definition}

We also define several classes of type distributions:

\begin{_definition}
	A type distribution $\td$ is called \emph{static} if $\forall t,\hat{t}\ \forall \theta \in \Theta : \td(t,\theta) = \td(\hat{t},\theta)$, else it is called \emph{dynamic}.
\end{_definition}

\begin{_definition}
	A type distribution $\td$ is called $pure$ if $\forall t\ \exists \theta \in \Theta : \td(t,\theta) = 1$, else it is called \emph{mixed}.
\end{_definition}

A SBG starts at time $t = 0$ in state $s^0$. In state $s^t$, the types $\theta_1^t,...,\theta_n^t$ are sampled from $\Theta$ with probability $\td(t,(\theta_1^t,...,\theta_n^t))$, and each player $i \in N$ is only informed about its own type $\theta_i^t$. Based on the history $H^t$, each player $i$ chooses an action $a_i^t \in A_i$ with probability $\pi_i(H^t,a_i^t,\theta_i^t)$. Given the joint action $a^t = (a^t_1,...,a^t_n)$, the game transitions into a successor state $s^{t+1}~\in~S$ with probability $T(s^t,a^t,s^{t+1})$ and every player $i$ receives an individual payoff given by $u_i(s^t,a^t,\theta_i^t)$. This process is repeated until the game reaches a terminal state $s^t \in \bar{S}$, after which the game stops.

Our definition of types follows the original definition of \cite{h1967}, which means that a type determines a player's payoffs and strategies. However, since we define strategies with respect to a \emph{history} of states and actions (rather than just the current state), a type may in fact specify strategies which change over time (such as players who learn or use recursive reasoning), and we thus also refer to it as \emph{behaviour}. Therefore, our interpretation of types is that of a ``programme'' which governs the behaviour of a player.

Each player may correspond to a specific role in the game. For instance, if we model a soccer team, player~1 may correspond to the goal keeper. Therefore, in the following sections, we implicitly assume that the ad hoc agent, denoted $\alpha$, controls the player of interest, denoted $i$, by which we mean that $\alpha$ chooses the strategy $\pi_i$. Furthermore, $i$ has a fixed type which is known to $\alpha$, and we denote its payoffs by $u_i(s^t,a^t,\alpha)$.

		% FLEXIBILITY & EFFICIENCY
		\subsection{Flexibility \& Efficiency}

Two important aspects of ad hoc coordination are \emph{flexibility} and \emph{efficiency}. We now define each of them formally within the SBG model. The definitions rely on the notion of \emph{paths} and probabilities of paths:

\begin{_definition}
	A \emph{path} $\rho$ in SBG $\Gamma$ is a sequence $\langle s_{\rho}^0,\theta_{\rho}^0,a_{\rho}^0,s_{\rho}^1,\theta_{\rho}^1,a_{\rho}^1,...,s_{\rho}^{t_{\rho}} \rangle$ where $s_{\rho}^\tau \in S$, $\theta_{\rho}^\tau \in \Theta$, $a_{\rho}^\tau \in A$, and $s_{\rho}^0 = s^0$. A path $\rho$ is \emph{terminating} if $s_{\rho}^{t_{\rho}} \in \bar{S}$, otherwise it is \emph{non-terminating}. Given a type distribution $\td$ for $\Gamma$, the probability of path $\rho$ is defined as $\pr (\rho | \Gamma, \td) = $
	\vspace{-5pt}
	\begin{equation*}
		\hspace{-1pt} \prod_{\tau = 0}^{t_{\rho}-1} \hspace{-2pt} \td(\tau,\theta_{\rho}^\tau) \, T(s_{\rho}^\tau,a_{\rho}^\tau,s_{\rho}^{\tau+1}) \hspace{-2pt} \prod_{k \in N} \hspace{-2pt} \pi_k(H_\rho^\tau,(a_{\rho}^\tau)_k,(\theta_{\rho}^\tau)_k)
	\end{equation*}
	where $H_\rho^\tau$ is the history extracted from $\rho$ until time $\tau$.
\end{_definition}

For $\pr (\rho | \Gamma, \td)$ to be well-defined (i.e. there is a set $X$ with $\forall \rho \in \hspace{-2pt} X \hspace{-2pt}:\hspace{-1pt} \pr (\rho | \Gamma, \td)\hspace{-1pt}\geq\hspace{-1pt}0$ and $\sum_{\rho \in X} \pr (\rho | \Gamma, \td)\hspace{-1pt}=\hspace{-1pt}1$), it is important to note the following two implications in the definition of SBGs. Firstly, no path $\rho$ can be prefixed by a terminating path, i.e., there is no $s_{\rho}^\tau \in \rho$ such that $\tau < t_{\rho}$ and $s_{\rho}^\tau \in \bar{S}$. This is important since otherwise $\pr (\rho | \Gamma, \td)$ might assign positive probability to a path which is prefixed by a terminating path and, thus, could never occur. Secondly, the only paths that can occur are either terminating (and hence finite) or non-terminating and \emph{infinite} (i.e. $t \rightarrow \infty$). Thus, if $\Phi$ is the set of all terminating paths and $\Psi$ the set of all infinite non-terminating paths, then $\sum_{\rho \in \Phi \cup \Psi} \pr (\rho | \Gamma, \td) = 1$.

Based on the notion of paths, we define the flexibility and efficiency of ad hoc agent $\alpha$ as follows:

\begin{_definition}
	Let $\Phi$ be the set of all terminating paths in SBG $\Gamma$. Given a set of type distributions $\TD$ for $\Gamma$, the \emph{flexibility} $F(\alpha | \Gamma, \TD)$ and \emph{efficiency} $E(\alpha | \Gamma, \TD)$ of $\alpha$ in $\Gamma$ with respect to $\TD$ are defined as
	\begin{eqnarray*}
		F(\alpha | \Gamma, \TD) & \hspace{-7pt} = \hspace{-7pt} & \frac{1}{|\TD|} \sum_{\td \in \TD} \sum_{\rho \in \Phi} \pr (\rho | \Gamma, \td) \\
		E(\alpha | \Gamma, \TD) & \hspace{-7pt} = \hspace{-7pt} & \frac{1}{|\TD|} \sum_{\td \in \TD} \sum_{\rho \in \Phi} \overline{\pr} (\rho | \Gamma, \td) \frac{ \left( \sum_{\tau = 0}^{t_{\rho}-1} u_i(s_{\rho}^\tau,a_{\rho}^\tau,\alpha) \right)^{r_1}}{(t_{\rho})^{r_2}}
	\end{eqnarray*}
	where $\overline{\pr} (\rho | \Gamma, \td) \hspace{-1pt}=\hspace{-1pt} \frac{\pr (\rho | \Gamma, \td)}{\sum_{\rho' \in \Phi} \pr (\rho' | \Gamma, \td)}$, and $r_1, r_2 \geq 1$ specify the relative importance between payoff and time.
\end{_definition}

$F(\alpha | \Gamma, \TD)$ and $E(\alpha | \Gamma, \TD)$ can be interpreted as, respectively, the average probability that $\alpha$ solves a task in $\Gamma$ and the average payoff per time step $\alpha$ received in solved tasks, where $\TD$ specifies all constellations of types that can occur. There may be problems in which flexibility is not a relevant metric because termination is guaranteed for some reason. In such cases, the primary metric is efficiency.

		% THE AD HOC COORDINATION PROBLEM
		\subsection{The Ad Hoc Coordination Problem}

We are now in a position to formally define the ad hoc coordination problem. The core aspect is that there is \emph{no prior coordination} between the ad hoc agent and the other agents in the system. We express this formally by requiring that the ad hoc agent does not know the type spaces $\Theta_j$ of the other players and, therefore, the type distribution $\td$ of the game.

\begin{_definition}
	Let $\Gamma$ be a SBG with type spaces $\Theta_j$, and let $\TD$ be a set of type distributions for $\Gamma$. The \emph{ad hoc coordination problem} is to optimise the flexibility $F(\alpha | \Gamma, \TD)$ and efficiency $E(\alpha | \Gamma, \TD)$ of ad hoc agent $\alpha$, subject to the constraint that $\alpha$ does not know $\Theta_j$ (and, therefore, the type distributions $\td$).
\end{_definition}

Computing $F(\alpha | \Gamma, \TD)$ and $E(\alpha | \Gamma, \TD)$ exactly is infeasible for all but the simplest games. We propose to approximate these by using the procedure given in Algorithm~\ref{alg:eval}. The procedure generates $K$ samples $F_k \hspace{-3pt} \sim \hspace{-3pt} F(\alpha | \Gamma, \TD)$ and $E_k \hspace{-3pt} \sim \hspace{-3pt} E(\alpha | \Gamma, \TD)$, based on which it approximates $F(\alpha | \Gamma, \TD) = \frac{1}{K} \sum_k F_k$ and $E(\alpha | \Gamma, \TD) = \frac{1}{K} \sum_k E_k$. Since all $F_k$ and $E_k$, respectively, come from the same distribution, by the law of large numbers this will converge to the true values of $F(\alpha | \Gamma, \TD)$ and $E(\alpha | \Gamma, \TD)$ for $K \rightarrow \infty$. The procedure needs some means to determine if a path is non-terminating. This could be done, for instance, by checking if the path reached a state space which contains no terminal states and cannot be left anymore, or by setting a maximum path length.

\begin{algorithm}[t]
	\linespread{1.20}
	\small
	\begin{algorithmic}
		\STATE \textbf{Input:} SBG $\Gamma$, set of type distributions $\TD$,
		\STATE \hspace{28pt} ad hoc agent $\alpha$, player $i$ (to be controlled by $\alpha$)
		\STATE \textbf{Output:} flexibility $F(\alpha | \Gamma, \TD)$, efficiency $E(\alpha | \Gamma, \TD)$
		\STATE $F \gets 0$
		\STATE $E \gets 0$
		\STATE \textbf{Repeat $K$ times:}
			\STATE \quad Randomly draw type distribution $\td \in \TD$
			\STATE \quad Generate path $\rho$ in $\Gamma$ with $\td$ ($\alpha$ controls $i$)
			\STATE \quad \textbf{If} $\rho$ terminates \textbf{do}
			\STATE \quad \quad $F \gets F + 1$
			\STATE \quad \quad $E \gets E + \left (\sum_{\tau=0}^{t_{\rho}-1} u_i(s_{\rho}^\tau,a_{\rho}^\tau,\alpha) \right)^{r_1} * (t_{\rho})^{-r_2}$
		\STATE $F(\alpha | \Gamma, \TD) \gets F / K$
		\STATE $E(\alpha | \Gamma, \TD) \gets E / K$
	\end{algorithmic}
	\caption{Evaluation procedure}
	\label{alg:eval}
\end{algorithm}

	% HARSANYI-BELLMAN AD HOC COORDINATION
	\section{Harsanyi-Bellman Ad Hoc Coordination} \label{sec:hba}

The problem of incomplete information is solved in Bayesian games by assuming that the type spaces $\Theta_j$ and type distribution $\td$ are common knowledge. This admits a solution in the form of the \emph{Bayesian Nash equilibrium} \cite{h1968a}, here defined for SBGs:

\begin{_definition}
	Let $H^t$ be the history at time $t$ and define $\Theta_{-i} = \times_{j \neq i} \, \Theta_{j}$. A \emph{Bayesian Nash equilibrium} (BNE) in state $s^t$ is a strategy profile $(\pi_1,...,\pi_n)$ in which, for all $i \in N$ and $\theta_i \in \Theta_i$, $\pi_i$ maximises
	\begin{equation}
		\sum_{\hat{\theta}_{-i} \in \Theta_{-i}} \hspace{-7pt} \td(t,\hat{\theta}_{-i} | \theta_i) \sum_{a \in A} u_i(s^t,a,\theta_i) \, \pi(H^t,a,(\theta_i,\hat{\theta}_{-i})) \label{eq:bne}
	\end{equation}
	where\\[-20pt]
	\begin{eqnarray*}
		\td(t,\theta_{-i} | \theta_i) & = & \frac{\td(t,(\theta_i,\theta_{-i}))}{\sum_{\hat{\theta}_{-i} \in \Theta_{-i}} \td(t,(\theta_i,\hat{\theta}_{-i}))} \\[5pt]
		\pi(H^t,a,\theta) & = & \prod_{k \in N} \pi_k(H^t,a_k,\theta_k).
	\end{eqnarray*}
\end{_definition}

In ad hoc coordination problems, the ad hoc agent does not know the type spaces $\Theta_j$ and, hence, the type distribution $\td$ of the game. Therefore, it cannot compute $\td(t,\theta_{-i} | \theta_i)$. However, using the history $H^t$, it can compute a \emph{posterior} $\pr (\theta_{-i} | H^t) = \prod_{j \neq i} \pr (\theta_j | H^t)$ with $\pr (\theta_j | H^t)$ being the probability that player $j$ has type $\theta_j$ based on history $H^t$
\begin{equation}
	\pr (\theta_j | H^t) = \frac{L(H^t | \theta_j) \, P(\theta_j)}{\sum_{\hat{\theta}_j \in \Theta_j} L(H^t | \hat{\theta}_j) \, P(\hat{\theta}_j)} \label{eq:post}
\end{equation}
where $L(H^t | \theta_j) = \prod_{\tau = 0}^{t-1} \pi_j(H^\tau,a_j^\tau,\theta_j)$ is the probability of history $H^t$ if the type of player $j$ is $\theta_j$, and $P(\theta_j)$ is the agent's prior belief that player $j$ has type $\theta_j$.

\hspace{-1pt}\cite{kl1993} studied single-state SBGs (with static pure type distributions) with players who choose actions to maximise their expected long-term payoff. They have shown that, if player $i$ maintains a posterior according to \eqref{eq:post}, and if the type distribution $\td$ is \emph{absolutely continuous} with respect to the posterior (i.e., $\td(t,(\theta_i,\theta_{-i})) > 0 \Rightarrow \pr (\theta_{-i} | H^t) > 0$), then player $i$'s predictions of future play will eventually be correct, regardless of player $i$'s own strategy (Theorem 1 in \cite{kl1993}). It follows that, if all players maintain such posteriors (where $\td$ is absolutely continuous with each posterior), and if all players choose their strategies according to a modified version of \eqref{eq:bne} which replaces the immediate payoff with the expected long-term payoff, then play will converge to a Nash equilibrium (NE) of the game (Theorem 2 in \cite{kl1993}). A similar result was shown by \cite{j1991} for myopic players (i.e. maximising immediate payoffs).

While these are encouraging theoretical results, there are several potential objections concerning the use of NE: Firstly, if there are multiple NE, then the players may converge to a sub-optimal equilibrium. Secondly, a NE is incomplete in that it does not specify strategies for off-equilibrium paths. Finally, \cite{dfl2004} have shown that if the posteriors of the players are not identical, then they might converge to a solution which is not a NE. However, our main concern with NE is that it makes strong behavioural assumptions about the players' behaviours (such as perfect rationality) which may be difficult to justify in ad hoc coordination. For instance, there is no guarantee that all players maintain posteriors according to \eqref{eq:post}. The same arguments hold for solution concepts in extensive form games, such as the perfect Bayesian equilibrium and sequential equilibrium \cite{ft1991}.

Rather than attempting to converge to NE, it is appealing to use \eqref{eq:bne} as a \emph{best-response} rule, since it maximises the expected payoff with respect to what types the ad hoc agent believes the other players to have and their strategies for all types. Based on Theorem 1 in \cite{kl1993}, we know that the agent's beliefs, and hence its expected payoffs, will be correct after some time. However, in its current form, \eqref{eq:bne} only considers immediate payoffs whereas optimal behaviour may require an agent to take payoffs of future states into account. Therefore, we propose to combine \eqref{eq:bne} with the Bellman optimality equation \cite{b1957} to obtain a best-response rule which we call \emph{Harsanyi-Bellman Ad Hoc Coordination}. Since ad hoc coordination requires that the agent does not know the type spaces $\Theta_j$, we assume instead that the ad hoc agent is provided with \emph{user-defined} type spaces $\Theta_j^*$, and we sometimes refer to $\Theta_j$ as the \emph{true} type spaces.

\begin{_definition}
	Let $\Gamma$ be an ad hoc coordination problem where ad hoc agent $\alpha$ controls player $i$ and has access to user-defined type spaces $\Theta_{-i}^* = \times_{j \neq i} \, \Theta_j^*$. \emph{Harsanyi-Bellman Ad Hoc Coordination} (HBA) is defined as $a_i^t \sim \arg\max_{a_i} E_{s^t}^{a_i}(H^t)$, where $E_s^{a_i}(\hat{H}) =$
	\begin{equation*}
		\hspace{-15pt} \sum_{\hspace{18pt} \theta^*_{-i} \,\in\, \Theta^*_{-i}} \hspace{-18pt} \pr (\theta^*_{-i} | H^t) \hspace{-19pt} \sum_{\hspace{20pt} a_{-i} \,\in\, A_{-i}} \hspace{-20pt} Q_s^{a_{i,-i}}( \hat{H} ) \, \prod_{j \neq i} \pi_j(\hat{H},a_j,\theta^*_j) \label{eq:hba}
	\end{equation*}
	is the expected long-term payoff for player $i$ of taking action $a_i$ in state $s$ after history $\hat{H}$ ($a_{i,-i} \triangleq (a_i,a_{-i})$), and $Q_s^a(\hat{H}) =$
	\begin{equation}
		\sum_{s' \in S} T(s,a,s') \left[ u_i(s,a,\alpha) + \gamma \max_{a_i}E_{s'}^{a_i} \hspace{-2pt} \left( \langle \hat{H},a,s' \rangle \right) \right] \label{eq:bellman}
	\end{equation}
	is the expected long-term payoff for player $i$ when joint action $a$ is executed in state $s$ after history $\hat{H}$, with $0 \leq \gamma \leq 1$ being the discount factor.
\end{_definition}

HBA is a modification of \eqref{eq:bne} which replaces $\td(t,\theta_{-i} | \theta_i)$ by the posterior $\pr (\theta_{-i} | H^t)$ \eqref{eq:post}, and in which the immediate payoff $u_i$ is replaced by an altered version \eqref{eq:bellman} of the Bellman optimality equation. The actual history $H^t$ is used to compute the posterior, and the projected histories $\hat{H}$ are used to generate all future trajectories.

Each user-defined type $\theta_j^* \in \Theta_j^*$ is a hypothesis about the behaviour of player $j$. While this gives HBA great flexibility (as $\Theta_j^*$ may include a variety of behaviours), it is important to note that the accuracy of \eqref{eq:bellman}, and hence efficiency of HBA, depends on how closely the user-defined types capture the players' true types. In this respect, we state two useful properties of HBA:

\begin{_proposition}
	Let $\Gamma$ be a SBG with static pure type distribution $\td$. If all players $i \in N$ are controlled by an HBA agent $\alpha_i$ with user-defined type spaces $\Theta_j^{*,i}$, and if $\forall j \neq i : \Theta_j \subseteq \Theta_j^{*,i}$, then play will converge to NE.
\end{_proposition}

This follows from Theorems 1 and 2 in \cite{kl1993} together with the fact that $\Theta_j \subseteq \Theta_j^{*,i}$ for all $i$ and $j$ (with $i \neq j$), which means that the type distribution $\td$ is always absolutely continuous with respect to the players' posteriors. Note that, while this proposition does not directly relate to ad hoc coordination, its does guarantee the minimum requirements of convergence and optimality in self-play, as formulated in \cite{bv2002}.

For the next proposition, we define the class of \emph{deterministic learners}, denoted $\Theta^D$, which consists of all types $\theta_j$ where, for all times $t$ and histories $H^t$, there exists a unique sequence $(\chi_{a_j})_{a_j \in A_j}$ such that $\pi_j(\langle H^t,(a,s) \rangle,a_j,\theta_j) + \chi_{a_j} = \pi_j(H^t,a_j,\theta_j)$, for all $(a,s) \in A \times S$. In other words, a deterministic learner always learns the same from a given history. By definition, this includes all fixed (i.e. non-changing) behaviours.

\begin{_proposition}
	Let $\Gamma$ be a SBG with static pure type distribution $\td$, where $\alpha$ controls $i$. If $\forall j \neq i : \Theta_j \subseteq \Theta^D \land \Theta_j \subseteq \Theta_j^*$, then $\alpha$ will be optimally efficient.
\end{_proposition}

This follows from the fact that there is some point after which HBA knows the players' types (Theorem 1 in \cite{kl1993}) and, since all types are deterministic learners, the expected payoffs \eqref{eq:bellman} are correct. Since HBA chooses actions with maximum expected payoffs, according to the Bellman principle \cite{b1957}, it follows that it achieves optimal efficiency. Note that HBA is itself a deterministic learner, hence HBA achieves optimal efficiency in self-play.

Both propositions assume that \eqref{eq:bellman} can be implemented directly, which is often infeasible. In Sections~\ref{sec:sim-exp} and \ref{sec:hum-exp}, we show how HBA can be implemented as a reinforcement learning procedure and an exact planning procedure.

		% TEMPORALLY REWEIGHTED POSTERIORS
		\subsection{Temporally Reweighted Posteriors} \label{sec:tpost}

A potential problem with the posterior defined in \eqref{eq:post} is that it assigns zero probability to a type $\theta_j$ if $\pi_j(H^t,a_j^t,\theta_j)$ is zero for any $t$. This can be problematic for the following reasons: If the game uses a dynamic or mixed type distribution, and if $\pr (\theta_j | H^t) = 0$ for a type $\theta_j$ that is not currently the true type of player $j$, then $\pr (\theta_j | H^\tau) = 0$ for all times $\tau > t$, even if player $j$'s type changes to $\theta_j$. Furthermore, if we have a user-defined type $\theta_j^*$ which approximates the true type $\theta_j$ of player $j$ in a subset $S^* \subset S$ (i.e. $\pi_j(H^t,a_j,\theta_j^*) \approx \pi_j(H^t,a_j,\theta_j)$ for $s^t \in S^*$), but not outside $S^*$, then \eqref{eq:post} might assign zero probability to $\theta_j^*$ once player $j$ leaves $S^*$. However, $\theta_j^*$ may be the best approximation we have for $S^*$, so it would be useful if \eqref{eq:post} was able to quickly reassign positive probability to $\theta_j^*$ once player $j$ returns to $S^*$. To address these problems, we introduce \emph{temporally reweighted posteriors}:

\begin{_definition}
	A \emph{temporally reweighted posterior} (TR-posterior) is defined as in \eqref{eq:post} by redefining
	\begin{equation}
		L(H^t | \theta_j) = \sum_{\tau=0}^{t-1} f(t-\tau) \, \pi_j(H^\tau,a_j^\tau,\theta_j) \label{eq:tpost}
	\end{equation}
	where $f(\xi) \geq 0$ and $f(\xi) \geq f(\xi+1)$, for all $\xi \in \mathbb{N}^+$.
\end{_definition}

The function $f$ is called the \emph{time weight} and can assume various forms. An example of a simple but useful time weight, called the \emph{general time weight}, is given by $f(\xi) = \max[ 0, a-b \hspace{1pt} (\xi-1)^c ]$ where $a,b,c \in \mathbb{R}_0^+$. This time weight can be used to produce various behaviours, depending on the parameters $a,b,c$. In particular, it can be used to give greater importance to more recent events, which means that HBA is able to quickly reassign probabilities. However, the crucial aspect of \eqref{eq:tpost} is that it defines a sum rather than a product, which means that the problems described above do not occur.

		% CONCEPTUAL TYPES
		\subsection{Conceptual Types} \label{sec:ctypes}

If the user-defined type space $\Theta_j^*$ for player $j$ does not include the true type space $\Theta_j$ (i.e. $\Theta_j \not\subset \Theta_j^*$), then $j$ might assume a type which is unknown to HBA, causing its expected payoffs to be inaccurate. In such cases, it would be useful if HBA was able to learn new types from experience. This opens up the possibility of using methods for opponent modelling (e.g. case-based reasoning \cite{wb2004} or recursive modelling \cite{gd2000}) which can be included in $\Theta_j^*$. In this work, we use a combination of case-based reasoning and fictitious play \cite{b1951}, called \emph{conceptual types}. Conceptual types are based on the observation that behaviour may not be specified on a state-by-state basis but rather on abstractions of state spaces. (An example are the ``information sets'' in extensive form games.) That is, there may be some \emph{world conceptualisation} inherent in a behaviour. While the types in $\Theta_j^*$ are used to hypothesise behaviours directly, a conceptual type can be used to hypothesise a world conceptualisation underlying a player's behaviour. Combined with the player's observed actions, this can be used to generalise actions to unseen states and increase accuracy in rarely visited states.

\begin{_definition}
	A \emph{conceptual type} (c-type) $\theta_j^c$ for player $j$ is a tuple $(d_j,r,f)$, where $d_j : S \times S \rightarrow \mathbb{R}_0^+$ is a symmetric distance function for pairs of states, $r \in \mathbb{R}^+$ is a radius, and $f$ is a time weight (as defined in Section~\ref{sec:tpost}), with
	\begin{equation*}
		\hspace{3pt} \pi_j(H^t,a_j,\theta_j^c) \hspace{-1pt} = \hspace{-2pt} \left\{ \hspace{-4pt} \begin{array}{rl} |A_j|^{-1} \hspace{1pt} \textbf{ if } \hspace{2pt} \hfill \nexists \tau \hspace{-2pt} < \hspace{-2pt} t \hspace{-1pt} : \hspace{-1pt} g(s^t,s^\tau) \hspace{-2pt} > \hspace{-2pt} 0 \hspace{1pt} \textbf{ else} \\[3pt] \eta \sum_{a^\tau \in H^t : \, a_j^\tau = a_j} f(t-\tau) \, g(s^t,s^\tau) \end{array} \right.
	\end{equation*}
	where $g(s_1,s_2) = \max\hspace{-2pt}\left[ 0 \, , 1 - d_j(s_1,s_2) \, r^{-1} \right]$ and $\eta$ is a normalisation constant s.t. $\sum_{a_j} \hspace{-1pt} \pi_j(H^t,a_j,\theta_j^c) = 1$.
\end{_definition}

The function $g$ is the hypothesised world conceptualisation of player $j$, where $d_j$ and $r$ specify how similar two states are from the perspective of player $j$ (examples given in Section~\ref{sec:sim-exp}). The time weight $f$ can be used to give greater importance to recent events, which allow c-types to adapt quickly to changing behaviours. Note that we can include multiple c-types in $\Theta_j^*$, each corresponding to a different world conceptualisation, and the posterior filters out those types which do not fit.

	% SIMULATED EXPERIMENTS
	\section{Simulated Experiments} \label{sec:sim-exp}

		% EXPERIMENTAL SETUP
		\subsection{Experimental Setup}

\begin{figure}[b]
	\centering
	\includegraphics[height=0.11\textheight]{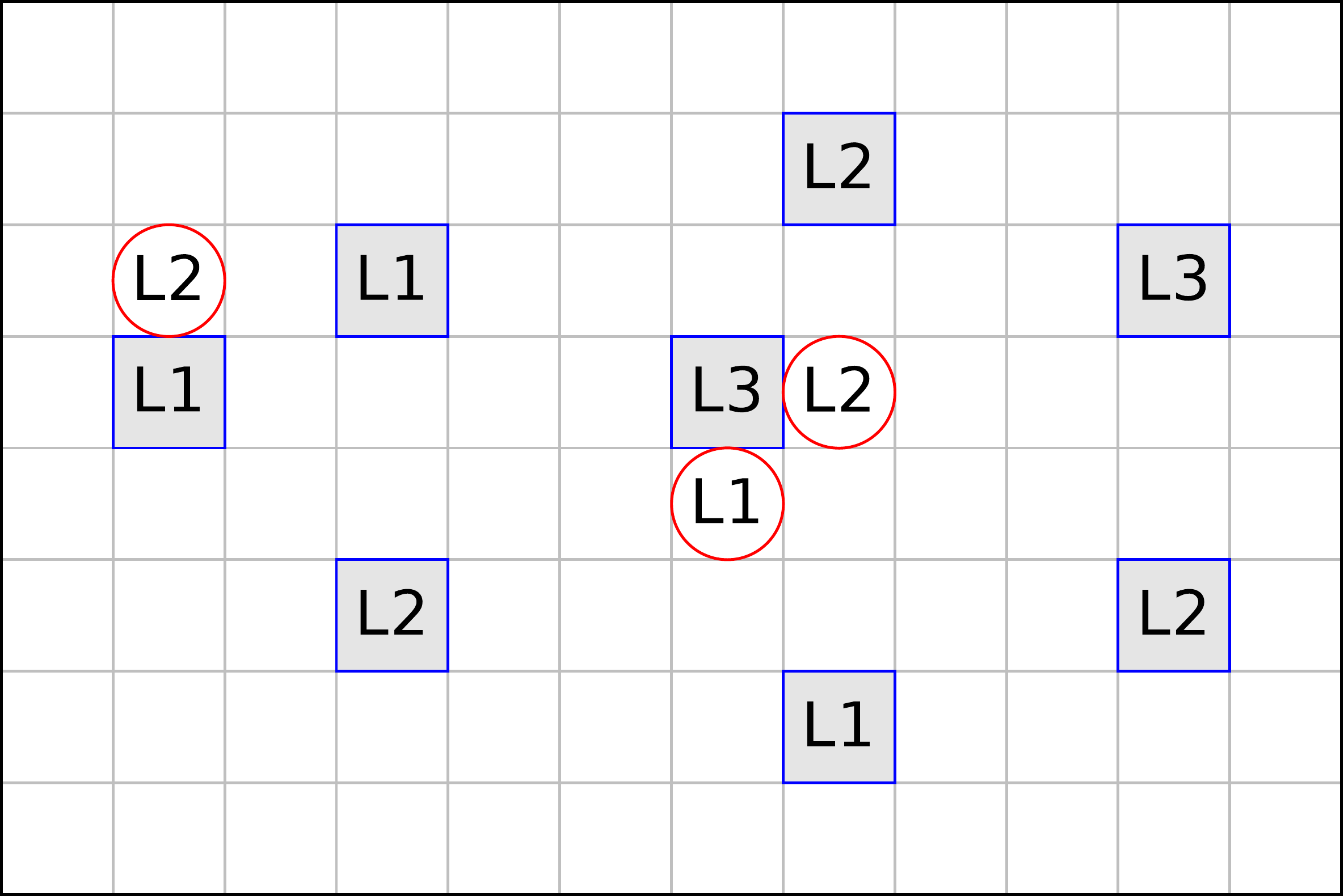}
	\hspace{5pt}
	\includegraphics[height=0.11\textheight]{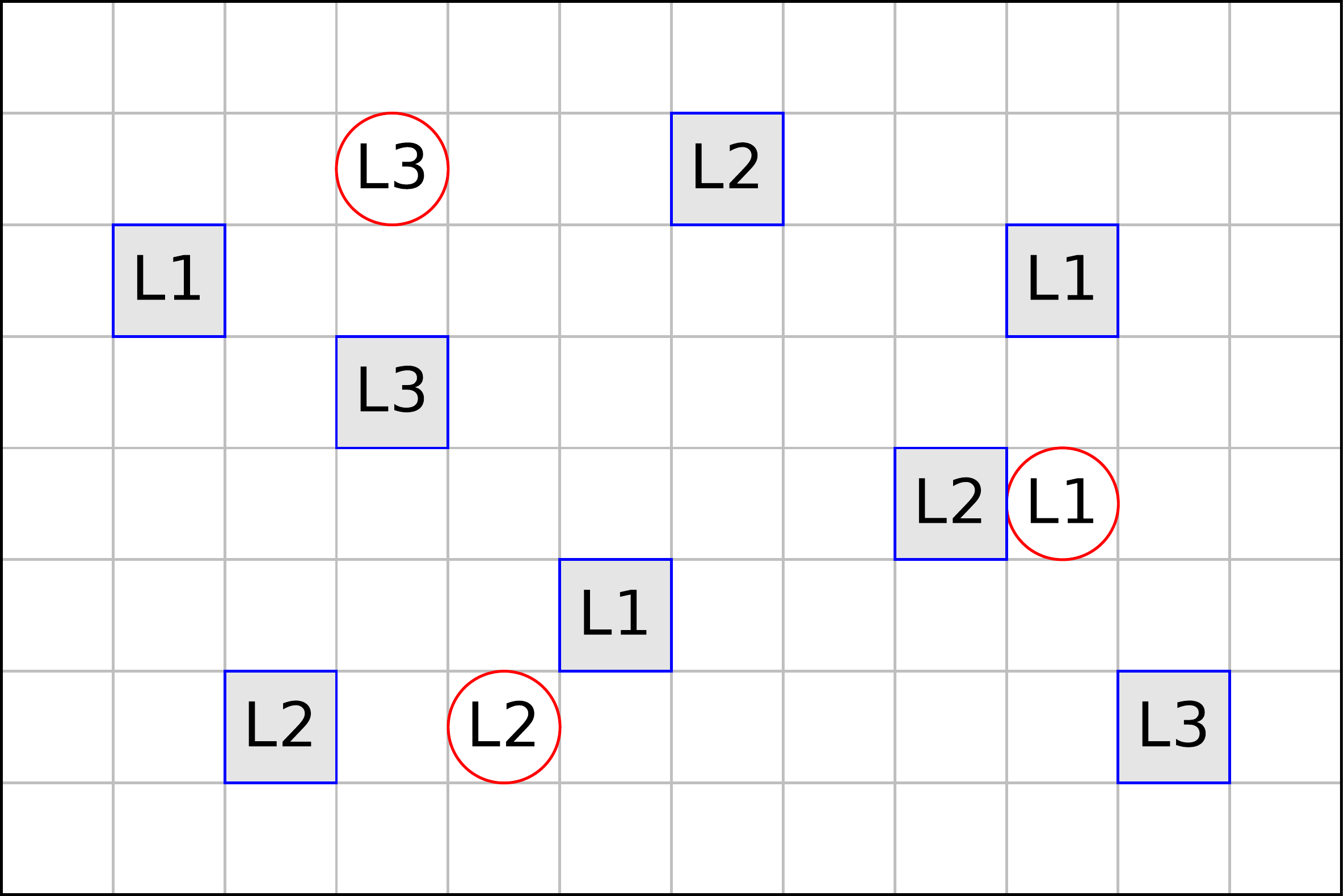}
	\caption{Level-based foraging domain. Players are marked by circles and foods are marked by squares (the levels are shown inside). \textit{Left:} Each player can load a food. \textit{Right:} No player can load a food.}
	\label{fig:foraging}
\end{figure}

We evaluated different configurations of HBA in a multiagent logistics domain called \emph{level-based foraging} (see Figure~\ref{fig:foraging}). A level-based foraging problem consists of a rectangular grid with $n$ players and $m$ foods. Each field in the grid is either empty or occupied by one player or one food. All players and foods have a \emph{level} ($\in \mathbb{N}^+$) where no food has a level greater than the sum of any 4 players' levels. A player can choose among 5 actions: $N$, $E$, $S$, $W$, and $load$. The first 4 actions move the player into the corresponding direction if the field is empty and inside the grid. A group of 1 to 4 players can \emph{load} a food if they are placed on fields next to the food and if the sum of their levels is at least as high as the food's level. A player which successfully loads a food obtains a payoff equal to the level of the loaded food. At all other times, it receives a negative payoff of -0.01. To avoid conflicts and keep this solvable, the foods are placed such that the Euclidean distance between each of them is greater than 1, and no food is placed at any border of the grid. The players' goal is to collect all foods in minimal time, while also trying to maximise their own payoffs. Since the players have different abilities (i.e. levels) and are spatially distributed, this requires strong coordination of their behaviours.

We specify 6 classes of types. The first 4 classes contain types with fixed behaviours (i.e. they do not change over time). They each have a parameter $\sigma$ which specifies the radius of their sight: H1 always goes to the closest visible food. H2 goes to the one visible food which is closest to the centre of all visible players. H3 always goes to the closest visible food with compatible level (i.e. it can load it) and H4 goes to the one visible food which is closest to all visible players such that the sum of their and H4's level is sufficient to load the food. H1-4 try to load the food once they are next to it. If they do not see a food, they go into a random direction. The last two classes specify types with learning behaviours: Class 5 contains all instances of JAL and class 6 all instances of CJAL, as specified in the next paragraph.

We evaluated various configurations of HBA and three alternative algorithms: JAL~\cite{cb1998} learns the action frequencies of each player in each state (i.e. opponent modelling) and uses them to compute expected action payoffs; CJAL~\cite{bs2007} is similar to JAL but learns the frequencies conditioned on its own actions; WoLF-PHC~\cite{bv2002} is a hill-climbing method in the space of mixed strategies. All three algorithms behave differently in ad hoc coordination \cite{ar2012}.

\begin{algorithm}[t]
	\linespread{1.20}
	\small
	\begin{algorithmic}
		\STATE Set $Q(s,a) \gets 0$ and $e(s,a) \gets 0$ for all $(s,a) \in S \times A$
		\STATE \textbf{Repeat} until $s^t \in \bar{S}$:
			\STATE \quad Observe: current state $s^t$
			\STATE \quad With probability $1 \hspace{-1pt} - \hspace{-1pt} \epsilon_1$: $a_i^t = \textsc{ChooseAction($s^t$)}$,
			\STATE \quad \hspace{95.5pt} else sample $a_i^t \sim A_i$
			\STATE \quad Observe: joint action $a^t$, own payoff $u_i^t$, next state $s^{t+1}$
			\STATE \quad \textsc{UpdateQ}($s^t,a^t,u_i^t,s^{t+1},e$)
			\STATE \quad \textbf{Repeat} $x$ times: \textsc{Expand}($d,s^{t+1},\textsc{Copy}(e)$)
		\vspace{5pt}
		\STATE \textsc{Expand}($d,s,\hat{e}$):
			\STATE \quad \textbf{Repeat} $d$ times or until $s \in \bar{S}$:
			\STATE \quad\quad With probability $1 \hspace{-1pt} - \hspace{-1pt} \epsilon_2$: $a_i = \textsc{ChooseAction($s$)}$,
			\STATE \quad\quad \hspace{95.5pt} else sample $a_i \sim A_i$
			\STATE \quad\quad $a_{-i} \gets \textsc{OppActions($s$)}$
			\STATE \quad\quad $(u_i,s') \gets \textsc{Simulate}(s,(a_i,a_{-i}))$
			\STATE \quad\quad \textsc{UpdateQ}($s,(a_i,a_{-i}),u_i,s',\hat{e}$)
			\STATE \quad\quad $s \gets s'$
		\vspace{5pt}
		\STATE \textsc{UpdateQ}($s,a,u,s',\hat{e}$):
			\STATE \quad $\delta = \beta ( u + \gamma \max_{\hat{a}_i} \hspace{-1pt} \textsc{ExpPay}(Q,s',\hat{a}_i) - Q(s,a) )$
			\STATE \quad $\hat{e}(s,a) \gets 1$
			\STATE \quad \textbf{For all} {$(\hat{s},\hat{a}) \in S \times A$ \textbf{s.t.} $\hat{e}(\hat{s},\hat{a}) \geq e_{min}$ \textbf{do:}}
				\STATE \quad\quad $Q(\hat{s},\hat{a}) \gets Q(\hat{s},\hat{a}) + \delta \, \hat{e}(\hat{s},\hat{a})$
				\STATE \quad\quad $\hat{e}(\hat{s},\hat{a}) \gets \lambda \, \hat{e}(\hat{s},\hat{a})$
		\vspace{5pt}
		\STATE \textsc{ChooseAction($s$)}:
			\STATE \quad \textbf{Return} $a_i \hspace{-1pt} \sim \hspace{-1pt} \arg\max_{\hat{a}_i} \hspace{-1pt} \textsc{ExpPay}(Q,s,\hat{a}_i)$
	\end{algorithmic}
	\caption{Reinforcement learning framework}
	\label{alg:exp1}
\end{algorithm}

A single framework (Algorithm~\ref{alg:exp1}) was used to implement each ad hoc agent. We assume that the ad hoc agent is able to observe the states of the game, each player's actions, and its own payoffs. For simplicity, we also assume that the agent knows the levels of all players and foods. The framework uses a table $Q$ to learn the expected long-term payoffs of joint actions, similar to Q-learning \cite{wd1992}. To accelerate learning, it uses an eligibility trace $e$ (see \cite{sb1998}) to connect current payoffs with past actions. We assume that the agent has access to a simulator \textsc{Simulate}($s,a$) which, based on the transition ($T$) and payoff ($u_i$) functions of the game, returns a successor state $s'$ and payoff $u$ after taking joint action $a$ in state $s$. This simulator is used in a sampling-based planning procedure \cite{kmn1999} \textsc{Expand}($d,s,\hat{e}$) which, starting in state $s$, generates a future trajectory of length $d$ and updates Q using the eligibility trace $\hat{e}$. The function \textsc{ExpPay}($Q,s,a_i$) computes the expected payoff for taking action $a_i$ in state $s$ based on $Q$, and the function \textsc{OppActions}($s$) samples actions for all other players $j \neq i$ in state $s$. HBA implements \textsc{Expand} using \eqref{eq:bne} and its posterior, and \textsc{OppActions} using its posterior and user-defined types. C/JAL implement these functions using their learned action frequencies. For WoLF-PHC, the framework defines $Q$ and $e$ on $S \times A_i$ (rather than $S \times A$) and \textsc{ExpPay}($Q,s,a_i$) is simply defined as $Q(s,a_i)$. Since WoLF-PHC does not model its opponents, we implement \textsc{OppActions} the same way as in JAL. The function \textsc{ChooseAction}($s$) is redefined to $a_i \sim \pi(s)$, where $\pi$ is the mixed strategy maintained in WoLF-PHC (cf. Tables 5 and 6 in \cite{bv2002}).

All algorithms used identical parameters: $\beta = .2$, $\gamma = .9$, $\lambda = .9$, $e_{min} = .01$, $\epsilon_1 = 0$, $\epsilon_2 = .2$, $x = 3$, $d = 20$. For WoLF-PHC, we used learning rates $\delta_w(t) = (1000 + \frac{t}{10})^{-1}$ and $\delta_l(t) = 2 \, \delta_w(t)$. For HBA, we used uniform prior beliefs ($P(\theta_j^*) = |\Theta_j^*|^{-1}$) and $a = 10$, $b = .01$, $c = 3$ for the general time weight. To obtain estimates of flexibility and efficiency, we used~Algorithm~\ref{alg:eval} with $i = 1$, $r_1 = r_2 = 1$, $K = 1000$, where we assumed a path to be non-terminating if it reached $t=1000$. The initial states were generated with random positions and levels for all players and foods, with the maximum level being equal to the number of players. All agents were tested on the same sequence of games and random numbers.

		% RESULTS
		\subsection{Results}

\begin{figure*}[t]
	\centering
	\includegraphics[height=0.12\textheight]{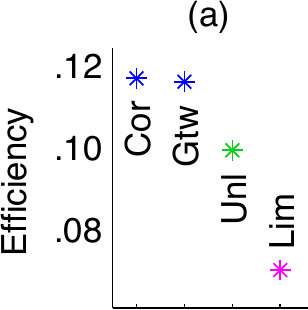}
	\hspace{10pt}
	\includegraphics[height=0.12\textheight]{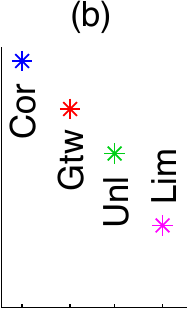}
	\hspace{10pt}
	\includegraphics[height=0.12\textheight]{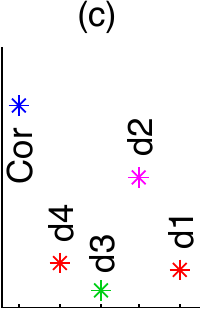}
	\hspace{30pt}
	\includegraphics[height=0.12\textheight]{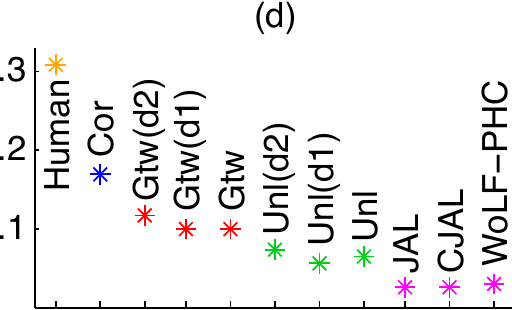}
	\caption{Results of simulated experiments, averaged over 1000 runs. Markers have the same colour if the difference is statistically insignificant (based on paired t-test with 5\% significance level). ``Cor'' is HBA with correct types, ``Gtw'' is HBA using TR-posterior with general time weight, ``Unl'' is HBA with unlimited normal posterior, and ``Lim'' is HBA with normal posterior limited to 9 most recent events.}
	\label{fig:exp1-res}
\end{figure*}

We tested the effectiveness of TR-posteriors by simulating the two situations described in Section~\ref{sec:tpost}. All tests were run on a $8 \times 8$ grid with 2 players and 5 foods. In Figure~\ref{fig:exp1-res}a, we used $\Theta_2 = \Theta_2^* = \left\{ \text{H1--H4} \, | \, \sigma = \infty \right\}$ and a dynamic pure type distribution which changed the type of player 2 after every 10 to 20 time steps. In Figure~\ref{fig:exp1-res}b, we used $\Theta_2 = \left\{ \text{H1--H4} \, | \, \sigma = 3,5,7 \right\}$, $\Theta_2^* = \left\{ \text{H1--H4} \, | \, \sigma = \infty \right\}$ (i.e the types in $\Theta_2^*$ were accurate only for subsets $S^* \subset S$) and a static pure type distribution. In both cases, the efficiency of HBA was significantly higher when using a TR-posterior with general time weight (Gtw) compared to both the normal posterior defined in \eqref{eq:post} (Unl) and a normal posterior which was limited to the 9 most recent events (Lim), which is the same time frame used in Gtw. In Figure~\ref{fig:exp1-res}a, Gtw even achieved the same efficiency as a version of HBA which always knew the correct type of the other player (Cor). All HBA agents achieved a perfect flexibility of 1.

We tested HBA with 4 conceptual types $\theta^c_j = (d^c_j,r,f)$ where $f(\xi) = [\xi < 10]_1$ and $r = 1$. In the following, we write $s.p_j$ ($s.f_k$) to refer to the position of player $j$ (food $f_k$) in state $s$, and $f_k \in s$ to say that food $f_k$ is available in state $s$. The distance functions $d^c_j$ are

$d_j^1(\hspace{-1pt}s_1,\hspace{-1pt}s_2\hspace{-1pt}) \hspace{-3pt} = \hspace{-3pt} [s_1 \neq s_2]_1 \, \infty$

$d_j^2(\hspace{-1pt}s_1,\hspace{-1pt}s_2\hspace{-1pt}) \hspace{-3pt} = \hspace{-3pt} [s_1.p_j = s_2.p_j \land \forall k : f_k \in s_1 \Leftrightarrow f_k \in s_2]_1 \, \infty$

$d_j^3(\hspace{-1pt}s_1,\hspace{-1pt}s_2\hspace{-1pt}) \hspace{-3pt} = \hspace{-3pt} \phi(s_1.p_j,\hspace{-1pt}s_2.p_j) \hspace{-1pt} + \hspace{-2pt} \sum_{k : f_k \in s_1 \underline{\lor} \hspace{1pt} f_k \in s_2} \hspace{-2pt} \psi(s_1.f_k,\mu)^{-\hspace{-1pt}\frac{3}{2}}$

$d_j^4(\hspace{-1pt}s_1,\hspace{-1pt}s_2\hspace{-1pt}) \hspace{-3pt} = \hspace{-3pt} d_j^3(s_1,s_2) \hspace{-2pt} + \hspace{-2pt} \sum_v \phi(s_1.p_v,s_2.p_v) \hspace{1pt} \omega_v^{-1.5}$

where $\phi(x_1,x_2) \hspace{-3pt} = \hspace{-3pt} \log(1 \hspace{-3pt} + \hspace{-3pt} \psi(x_1, \hspace{-1pt} x_2)) \frac{1}{2}$, $\mu \hspace{-2pt} = \hspace{-3pt} s_1.p_j \hspace{-3pt} + \hspace{-3pt} \frac{1}{2} (s_2.p_j \hspace{-3pt} - \hspace{-3pt} s_1.p_j)$, $\omega_v \hspace{-2pt} = \hspace{-2pt} \min[ \psi(s_1.p_v,\mu),\psi(s_2.p_v,\mu) ]$, and $\psi(x_1,x_2)$ denotes the Euclidean distance between $x_1$ and $x_2$. All tests were run on a $8 \times 8$ grid with 2 players and 5 foods, using $\Theta_2 = \left\{ \text{H1--H4,JAL,CJAL} \, | \, \sigma = \infty \right\}$ (C/JAL used same parameters as HBA), $\Theta_2^* = \left\{ \theta_2^c \right\}$ (each $c=1,...,4$ tested separately), and a static pure type distribution. The results in Figure~\ref{fig:exp1-res}c show that HBA achieved good efficiency (compared to Cor) using $\theta_j^2$, while the other c-types were less efficient. All HBA agents achieved statistically equivalent flexibilities of $0.86 \pm 0.01$.

Finally, we tested HBA, JAL, CJAL, and WoLF-PHC on a $10 \times 10$ grid with 3 players and 8 foods, using $\Theta_{2,3} = \left\{ \text{H1--H4,JAL,CJAL} \, | \, \sigma = 5,7,9 \right\}$ and $\Theta_{2,3}^* = \left\{ \text{H1--H4} \, | \, \sigma = \infty \right\}$. To add more realism, players 2 and 3 were ``defective'' with probability 0.2, where a defective player changed its type randomly every 10 to 30 time steps. While the potential of HBA is demonstrated by Cor, it would also be useful to know the optimal solution to the problem. However, with a complex problem such as this one, we were unable to compute optimal solutions. Instead, we had 6 \emph{humans} play the game in a graphical user interface (each one played the full 1000 runs, distributed over 7 days at their own convenience), where no human was familiar with the technical details of this work. We do not necessarily claim that humans produce optimal solutions, but we expect them to perform consistently well in this setting. To cope with the increased problem size, we set the planning power of the algorithms to $x = 10$ and $d = 30$ (cf. Algorithm~\ref{alg:exp1}).

The results (Figure~\ref{fig:exp1-res}d) show that HBA clearly outperformed all alternative algorithms, with Unl and Gtw being over 100\% and 200\% more efficient, respectively. This is despite the fact that the user-defined types $\Theta_{2,3}^*$ did not include any true types of the players. We also tested HBA with the c-types $\theta_j^1$ and $\theta_j^3$ (added separately to $\Theta_{2,3}^*$) but found that the efficiency of HBA did not improve significantly. This is since C/JAL learned similar behaviours to H1 and H3, which were already covered in $\Theta_{2,3}^*$. We found that HBA's posteriors often assigned high probabilities to H1/3 when the true type of the player was in fact C/JAL. Since H1/3 ignore other players, this means that C/JAL did not effectively coordinate their behaviours with other players. We found similar results for WoLF-PHC. As was expected, the humans achieved high efficiency (Figure~\ref{fig:exp1-res}d shows the best human) and outperformed even Cor. One reason for this is the fact that the humans had much greater planning power than HBA. Lastly, HBA achieved higher flexibilities ($.83 \pm .01$) than JAL ($.734$), CJAL ($.749$), and WoLF-PHC ($.744$), while the humans all achieved perfect flexibility ($1.0$).

	% HUMAN-MACHINE EXPERIMENT
	\section{Human-Machine Experiment} \label{sec:hum-exp}

		% EXPERIMENTAL SETUP
		\subsection{Experimental Setup}

We conducted a large-scale human-machine experiment at the Royal Society Summer Science Exhibition 2012. Therein, the human participants played repeated \emph{Prisoner's Dilemma} (PD) and \emph{Rock-Paper-Scissors} (RPS) against HBA and alternative algorithms, where each game was played for 20 rounds. We collected data from 427 participants, of which 186 played PD and 241 played RPS. The lowest and highest recorded ages were 9 and 72, respectively, with an average age of about 17.

A large public exhibition such as this one is an excellent testbed environment for ad hoc agents, since the visitors vary widely in factors such as age, intelligence, and behaviour. However, in order to make statistically relevant comparisons, we required data from many participants. Therefore, the games needed to be simple enough so participants would understand them quickly, yet they also needed to be interesting in terms of coordination strategies. PD and RPS are two widely studied problems in game theory which we believe cover these properties. In PD, the symmetric payoffs are $u_1(C,C) \hspace{-2pt} = \hspace{-2pt} 3$, $u_1(D,D) \hspace{-2pt} = \hspace{-2pt} 1$, $u_1(C,D) \hspace{-2pt} = \hspace{-2pt} 0$, $u_1(D,C) \hspace{-2pt} = \hspace{-2pt} 5$. The problem here is that the only NE, and hence stable outcome, is at (D,D), while (C,C) is the only outcome that has both the highest welfare (sum of payoffs) and fairness (product of payoffs) but is unstable since the players could deviate to obtain higher immediate payoffs. In RPS, the payoffs are +1/0/-1 for won/even/lost games. The only NE is for all players to play randomly. However, even if humans attempt to play randomly, they often fall back to patterns \cite{w1972} against which the other player can coordinate its actions.

Our hypothesis for the experiment was that the human would switch between several simple behaviours, as opposed to having one complex behaviour. Therefore, we modelled the problem as a SBG with a dynamic mixed type distribution (unknown to us) which governed the type of the human, and we provided HBA with a small set of types (given in Tables~\ref{tab:pd-types} and \ref{tab:rps-types}) which we believed the human could have. HBA did not use any conceptual types.

\begin{algorithm}[t]
	\linespread{1.20}
	\small
	\begin{algorithmic}
		\STATE \textbf{Repeat:}
			\STATE \ \ Observe current state $s^t$
			\STATE \ \ \textbf{For all} $a_i \in A_i$ \textbf{do}:
				\begin{eqnarray*}
					\Omega(a_i) & \hspace{-5pt}=\hspace{-5pt} & \left\{ \langle s^t,a^t,...,s^{t+l},a^{t+l} \rangle \, | \, a_i^t = a_i \right\} \\
					 & & \text{where } l = \min[ l^*,t^* - t] - 1 \\
					E(a_i) & \hspace{-7pt}=\hspace{-8pt} & \sum_{\omega \in \Omega(a_i)} \left[ \prod_{\tau=t}^{t+l} \textsc{OppStrat}(s^\tau,a^\tau) \sum_{\tau=t}^{t+l} u_i(s^\tau,a^\tau) \right]
				\end{eqnarray*}
			\STATE \ \ Sample action $a_i^t \sim \arg\max_{a_i} E(a_i)$
	\end{algorithmic}
	\caption{Exact planning framework}
	\label{alg:exp2}
\end{algorithm}

The alternative algorithms were CJAL for PD, which was shown to outperform both JAL and WoLF-PHC in PD \cite{bs2007}, and JAL for RPS, which is guaranteed to converge to NE in self-play in zero-sum games \cite{b1951}. We implemented all algorithms using a single framework (Algorithm~\ref{alg:exp2}), where we set $l^* = 10$ for PD, $l^* = 1$ for RPS, and $t^* = 20$. The function \textsc{OppStrat}($s^\tau,a^\tau$) returns the probability that players $j \neq i$ choose actions $a_j^\tau$ in state $s^\tau$. HBA implements this by averaging over all user-defined types in $\Theta_j^*$ using its current posterior, and C/JAL do this using their learned actions frequencies. While PD and RPS have no states, we found that the performance of C/JAL could be further improved by introducing ``artificial'' states, which we simply defined as $s^t = a^{t-1}$ (in the first round, C/JAL assumed the opponent to play randomly). HBA used uniform prior beliefs and the general time weight with $a = 10$, $b = 0.05$, $c = 3$.

The procedure of the experiment was as follows: First, we randomly sampled a participant from the set of visitors which were currently at our exhibit. The participant was then brought to a dedicated table with a chair and a laptop on it. The laptop ran a programme, with an intuitive graphical user interface, which prompted the participant to choose between PD and RPS. The rules of the games were explained both textually in the programme and in person by one of our staff members to make sure the participant understood the rules. The game was then played in two matches, each lasting 20 rounds. One of the matches was against HBA and the other match against C/JAL, but this was hidden from the participant and the order was chosen randomly. The programme displayed the current match, round, and scores of all players, and also allowed to display the rules at any time. At the end of each round, the participant was shown the actions and scores of both players, and at the end of each match, the participant was given a summary of the scores.

\begin{figure*}[t]
	\centering
	\includegraphics[height=0.13\textheight]{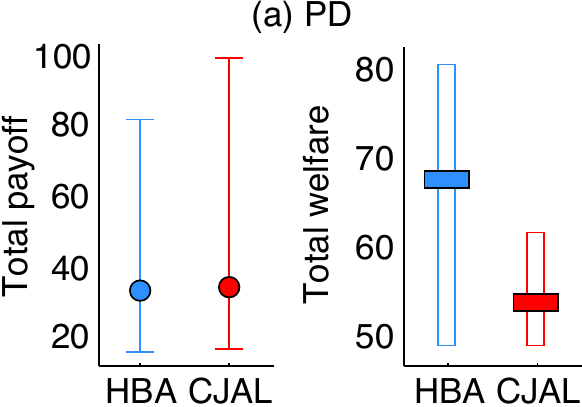}
	\hspace{30pt}
	\includegraphics[height=0.13\textheight]{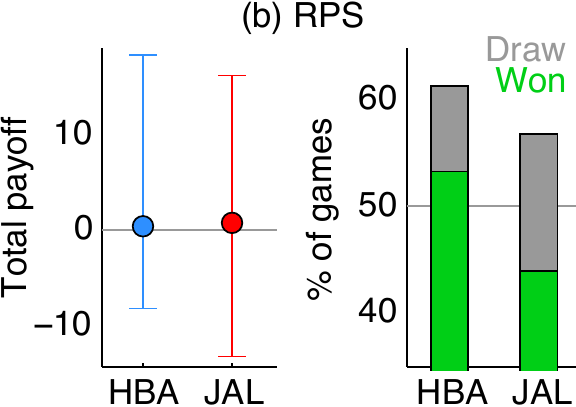}
	\hspace{30pt}
	\includegraphics[height=0.13\textheight]{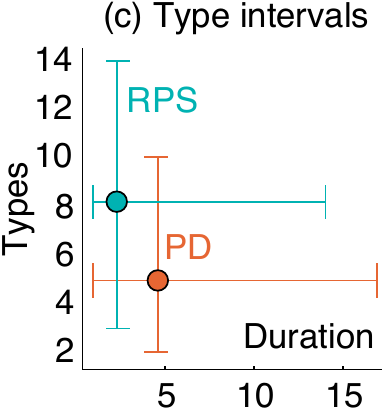}
	\caption{Results of the human-machine experiment. Circles and whiskers correspond to mean, minimum, and maximum values, respectively. The welfare plot in (a) shows the median value and 25\%/75\% percentiles.}
	\label{fig:exp2-results}
\end{figure*}

		% RESULTS
		\subsection{Results}

In the following, all significance statements are based on paired t-tests with 5\% significance level. Figures~\ref{fig:exp2-results}a and \ref{fig:exp2-results}b show the results for PD and RPS, respectively. In both games, the average total payoffs of HBA and C/JAL were statistically equivalent. Since the time was fixed to 20 rounds, it means that they achieved equal efficiency. This is, in fact, a positive result considering that C/JAL are strong candidates in PD/RPS. In addition, as we discuss in the following, HBA behaved very differently from C/JAL, with beneficial side effects.

In PD, the most desirable long-term outcome is (C,C) since it is both welfare and fairness optimal, and since it is a non-myopic equilibrium \cite{b1993}, meaning that no player has a long-term incentive to deviate. With this in mind, we point out that in over 28\% of the games, HBA and the human played (C,C) in at least 50\% of the final 10 rounds of the game, while CJAL did not achieve this in any game. Thus, HBA achieved a significantly higher total welfare than CJAL (Figure~\ref{fig:exp2-results}a). This is despite the fact that neither of them was optimised for social welfare. The reason for this is that HBA was planning more accurately than CJAL. When computing the expected payoffs $E(a_i)$, CJAL uses its learned action frequencies to obtain probabilities for each trajectory in $\Omega(a_i)$. However, these probabilities can only be accurate for states that have been visited frequently enough. Moreover, if a player changes its behaviour, CJAL requires new evidence from all states to accurately reflect the change. On the other hand, HBA uses its posterior and types to compute probabilities of trajectories. Therefore, once HBA has an accurate posterior, it can use the types to accurately plan in the entire state space of the game, including unseen states. This also allows HBA to plan the effects of its actions on the other player, which means that HBA may take actions to manipulate the player's decisions. Finally, if a player changes its behaviour, HBA only needs to update its posterior, which requires much less information than the update in CJAL.

In RPS, the crucial questions is whether a player is winning or not. Interestingly, the winning rate of HBA (53.71\%) was significantly higher than the winning rate of JAL (43.98\%), as shown in Figure~\ref{fig:exp2-results}b. While in PD the good performance of HBA was due to its planning capabilities, in RPS this was not as relevant since the planning horizon was limited to trajectories of length~1. Rather, HBA's good performance was due to the fact that it recognised changed behaviours faster than JAL. Indeed, in a game such as RPS, it can be expected that the human players change frequently between different strategies. This is confirmed by the statistics shown in Figure \ref{fig:exp2-results}c, which show the average number of types used by the human players and the average duration. The statistics are based on HBA's posteriors, where the number of types for player $i$ in a play corresponds to the number $q$ in $\langle t_0,t_1,...,t_q \rangle$, with $t_0 = 0$ and $t_q = 20$, for which $\arg\max_{\theta_i} \pr (\theta_i | H^\tau) \subseteq \arg\max_{\theta_i} \pr (\theta_i | H^{\tau+1})$ for all $t_{y-1} \leq \tau < t_y$ and $y \in \left\{ 1,...,q \right\}$, and where the average duration is $\frac{1}{q} \sum_y t_y-t_{y-1}$. According to these statistics, the human players had 4.45 types with a duration of 4.96 rounds in PD, and 8.25 types with a duration of 2.46 rounds in RPS. Clearly, with a duration of only 2.46 rounds, planning was not as important as recognising changed types. By using TR-posteriors, HBA was able to do this effectively.

	% CONCLUSION
	\section{Summary \& Open Questions} \label{sec:conc}

This work is concerned with the ad hoc coordination problem, in which the goal is to design an autonomous agent (the ad hoc agent) which can achieve optimal flexibility and efficiency in a multiagent system in which the behaviour of the other agents is not a priori known. We make three important contributions to the ad hoc coordination problem:

\vspace{-5pt}

\begin{enumerate}
	\item We propose a game-theoretic model, SBG, which captures the notion of private information in the form of types. Based in this model, we give formally concise definitions of flexibility, efficiency, and the ad hoc coordination problem. We also provide a procedure which can be used to estimate the ad hoc agent's flexibility and efficiency.
	\item From this model, we derive a principled solution, HBA, which utilises a set of user-defined types in a planning procedure to find optimal actions in the sense of Bayesian Nash equilibrium and Bellman optimal control. We also propose two possible extensions which enable HBA to recognise changed types and learn new types.
	\item We show how HBA can be implemented as a reinforcement learning and exact planning procedure, and we provide extensive empirical evaluations in a complex multiagent logistics domain and a large-scale human-machine experiment. Our results show that HBA is both more flexible and efficient than alternative methods.
\end{enumerate}

The work presented in this paper provides a rich ground for future research, including the following open questions:

\vspace{-5pt}

\begin{itemize}
	\item A crucial design parameter of HBA are the user-defined type spaces $\Theta_j^*$ provided to it. In this regard, an important direction for future research would be to analyse how closely $\Theta_j^*$ must approximate $\Theta_j$ for HBA to be able to achieve optimal flexibility and efficiency.
	\item Another design parameter of HBA is the posterior $\pr(\cdot | H^t)$, and in this work we discussed two different formulations (the product posterior and TR-posteriors). It would be interesting to explore alternative posterior formulations and to analyse the conditions under which they are guaranteed to converge to the type distribution of the game.
	\item The prior belief $P$ can be considered a meta-parameter of HBA (it is a parameter of the posterior, which in turn is a parameter of HBA), and in our experiments we assumed that the prior beliefs were uniform. An interesting question in this regard is whether HBA could automatically derive prior beliefs from the user-defined type spaces so as to further maximise its efficiency.
	\item HBA currently assumes that an expert can provide manually specified types for the problem at hand. However, this can be a cumbersome task in complex domains. Future work could investigate how HBA might generate useful types from the problem description so that the burden of having to manually specify types can be alleviated, or perhaps eliminated altogether.
	\item Finally, as we employ HBA in increasingly complex problem domains, it becomes apparent that the type specifications, likewise, become increasingly complex. One way to reduce this type complexity might be to use a hierarchical type specification, in which types are structured into smaller sub-types.
\end{itemize}

{\renewcommand{\arraystretch}{1.5}
\begin{table*}[t]
	\centering
	\begin{tabular}{|l|l|}
		\hline
		\textbf{PD type} & \textbf{Definition} \\
		\hline
		AlwaysC & $a_i^t = C$ \\
		\hline
		TitForTat & $a_i^0 = C, a_i^t = a_j^{t-1}$ \\
		\hline
		TitFor2Tats & $a_i^{0,1} = C, a_i^t = C \textbf{ if } a_j^{t-1,t-2} = C \textbf{ else } D$ \\
		\hline
		Optimistic & $\pi_i(C,H^t) = 1 \textbf{ if } t < 2 \lor a_j^{t-1} = C \lor \mu = 0 \textbf{ else } 0.2 + 0.8 \sigma$ \\
		\hline
		Pessimistic & $\pi_i(D,H^t) = 1 \textbf{ if } t < 2 \lor a_j^{t-1} = D \textbf{ else } 0.2 + [\mu > 0]_1 0.8 \sigma$ \\
		\hline
		& $\mu = \sum_{\tau=0}^{t-2} [a_i^\tau=C]_1, \, \sigma = \frac{1}{\mu}\sum_{\tau=0}^{t-2} [a_i^\tau=a_j^{\tau+1}=C]_1$ \\
		\hline
	\end{tabular}
	\caption{PD types. $[b]_1 = 1$ iff. $b$ is true, else 0.}
	\label{tab:pd-types}
\end{table*}}

{\renewcommand{\arraystretch}{1.5}
\begin{table*}[t]
	\vspace{10pt}
	\centering
	\begin{tabular}{|l|l|}
		\hline
		\textbf{RPS type} & \textbf{Definition} \\
		\hline
		Copycat & $a_i^0 \sim U(A_i), a_i^t = a_j^{t-1}$ \\
		\hline
		RetryIfWon & $a_i^t \sim U(A_i) \textbf{ if } t = 0 \lor u_i(a^{t-1}) < 0 \textbf{ else } a_i^t = a_i^{t-1}$ \\
		\hline
		$i$-focused($h$) & $\pi_i(a_i,H^t) =  g(a_i,x) / \sum_{\hat{a}_i \in A_i} g(\hat{a}_i,x), \, x = \min[t,h]$ \\
		\hline
		$h \in \left\{ 1,2 \right\}$ & $g(a_i,x) = \max\hspace{-2pt}\left[ 0,x - \sum_{\tau=1}^x [a_i^{t-\tau}=a_i]_1 (x+1-\tau) \right]$ \\
		\hline
		$j$-focused($h$) & $a_i^t \sim \arg\max_{a_i} \sum_{a_j \in A_j} \pi_j(a_j,H^t) \, u_i(a_i,a_j)$ \\
		\hline
		$h \in \left\{ 1,2 \right\}$ & where $\pi_j(a_j,H^t)$ is obtained using $i$-focused($h$) for $j$ \\
		\hline
	\end{tabular}
	\caption{RPS types. $U$ is the uniform distribution.}
	\label{tab:rps-types}
\end{table*}}

	%ACKNOWLEDGEMENTS
	\section*{Acknowledgements}

This work was partially supported by grants from the UK Engineering and Physical Sciences Research Council (EP/H012338/1), the European Commission (TOMSY Grant 270436, FP7-ICT-2009.2.1 Call 6) and a Royal Academy of Engineering Ingenious grant.

	% BIBLIOGRAPHY
	\bibliographystyle{apalike}
	\bibliography{uai13_adhoc}

\end{document}